\documentclass[pre,twocolumn,nobalancelastpage,amsmath,amssymb,showpacs,
nofootinbib, superscriptaddress]{revtex4-2}

\usepackage{graphics}      
\usepackage{graphicx}      
\usepackage{longtable}     
\usepackage{url}           
\usepackage{bm}            
\usepackage{amsmath}
\usepackage{dcolumn}
\usepackage{mathtools}
\usepackage{hyperref}
\usepackage[dvipsnames]{xcolor}

\newcommand{\gdot}{\dot\gamma}

\newcommand{\edot}{\dot\epsilon}

\begin{document}
\title{Critical Scaling of Compression-Driven Jamming of {\color{black}Athermal} Frictionless Spheres {\color{black}  in Suspension}}
\author{Anton Peshkov}
\affiliation{Department of Physics and Astronomy, University of Rochester, Rochester, NY 14627}
\author{S. Teitel}
\affiliation{Department of Physics and Astronomy, University of Rochester, Rochester, NY 14627}
\date{\today}

\begin{abstract}
{\color{black}We numericaly study a system of athermal, overdamped, frictionless spheres, as in a non-Brownian suspension, in two and three dimensions.  Compressing the system  isotropically at a fixed rate $\dot\epsilon$, we investigate the critical behavior at the jamming transition.} The  finite compression rate introduces a control time scale, which allows one to probe the critical time scale associated with jamming.  As was found previously for steady-state shear-driven jamming, we find for compression-driven jamming that pressure obeys a critical scaling relation as a function of packing fraction $\phi$ and  compression rate $\dot\epsilon$, and that the bulk viscosity $p/\dot\epsilon$ diverges upon jamming. A scaling analysis  determines the critical exponents associated with the compression-driven jamming transition. Our results suggest that  stress-isotropic, compression-driven, jamming may be in the same universality class as  stress-anisotropic, shear-driven, jamming.
\end{abstract}
\maketitle


Athermal granular and related soft matter materials, such as non-Brownian suspensions, emulsions, and foams, all undergo a phase transition from a liquid-like state to a rigid  disordered  state as the packing fraction $\phi$ of the granular particles increases.  This is  the jamming transition \cite{LiuNagel,OHern}.  Here we focus on the behavior of frictionless particles, where jamming is like a continuous phase transition {\color{black}with respect to the behavior of the stress}.  Early studies of jamming focused on what we will call stress-isotropic jamming: mechanically stable jammed configurations are generated by  isotropically compressing the system, or by energy quenching random initial configurations at fixed $\phi$ \cite{OHern,Wyart,Chaudhuri,Ciamarra,Vagberg.PRE.2011}.  At low $\phi$ particles  avoid each other and the  pressure $p$ vanishes.  At a critical $\phi_J$ a system spanning rigid cluster forms and the  pressure becomes finite, while the shear stress $\sigma$ remains zero.  
Later studies investigated
shear-driven jamming \cite{OT1,OT2,VagbergOlssonTeitel,OT3, Hatano1,Hatano2,Hatano3,Otsuki,Heussinger1,Heussinger2}, where  the system is uniformly sheared at a fixed strain rate $\dot\gamma$.  For systems with a Newtonian rheology, such as  particles in suspension, the system flows at low $\phi$ and small $\dot\gamma$ with a shear stress $\sigma\propto \dot\gamma$.  
Thus, for $\dot\gamma\to 0$, the viscosity $\eta=\sigma/\dot\gamma$ remains finite. However, above a critical $\phi_J$, the system develops a non-zero yield stress $\sigma_0(\phi)=\lim_{\dot\gamma\to 0}\sigma>0$ leading to a diverging viscosity. 
Because of this finite $\sigma$, we will refer to this as stress-anisotropic jamming. Given the different symmetry of  anisotropic shear-driven jamming vs  isotropic compression-driven jamming, it is natural to wonder if they belong to the same critical universality class, 
{\color{black}i.e., if the critical exponents describing  singular behaviors  are the same {\color{black} for any given dimensionality of the system}. For equilibrium critical points, different symmetries often imply different universality classes \cite{CL}.

In this work we consider this question {\color{black}by investigating the  dynamical behavior of the unjammed state below $\phi_J$, in order to probe the diverging time scale associated with jamming.  In particular, we}
numerically compute the bulk viscosity $\zeta=p/\edot$ of frictionless, overdamped, soft-core particles, isotropically compressed at finite compression rates $\edot$.  
{\color{black}Although isotropic compression causes the packing $\phi$ to steadily increase, and thus it does not produce a steady-state ensemble as does simple shearing, we nevertheless can compute $\zeta$ by averaging results over  several different independent compression runs.}
Below jamming we find that $\zeta$ has a well defined limit as $\edot\to 0$, which diverges as $\phi\to\phi_J$.
We  demonstrate that a  simple critical scaling ansatz, found previously to apply for shear-driven jamming \cite{OT2,VagbergOlssonTeitel}, also applies to compression-driven jamming, thus uniting these two different thrusts of jamming research and providing a framework in which to numerically address the question of a common universality class.
Our  scaling analysis  strongly suggests that the critical exponents of compression-driven jamming in two dimensions (2D) are the same as  previously found for shear-driven jamming; the situation in three dimensions (3D) remains less clear.
}

{\color{black}{\bf Prior Works}: 
Numerical works in 3D \cite{Baity,Jin2} have argued for a common universality for athermal isotropic  and anisotropic jamming, by looking at static ``shear-jammed" configurations of soft-core spheres, obtained by applying a static shear strain to unjammed isotropic configurations, and increasing the shear strain until jamming occurs.  The same scalings of pressure and contact number were obtained as were previously found in isotropic jamming \cite{OHern}.  Similar conclusions for {\em thermalized} hard-core spheres have been found in infinite-dimensional mean-field calculations \cite{Urbani} and in 3D simulations \cite{Jin2}.
These works are concerned with the  structural properties of static, mechanically stable, configurations at or above jamming, and  do not probe the dynamics associated with a diverging time scale as one approaches jamming from below.

However, a connection between structural and dynamic properties was proposed in \cite{DeGiuli,During} using a marginal-stability analysis.  If $\eta_p=p/\gdot$ is the pressure analog of shear viscosity in a shear-driven steady state, then \cite{DeGiuli,During} argued that the exponent $\beta$, which characterizes the divergence of $\eta_p$ as jamming is approached from below, is  determined by the exponent $\theta$ that describes the distribution of small contact forces between particles in configurations exactly at jamming.  In other works \cite{Olsson3D,OlssonRelax}, this viscosity $\eta_p$ was found  to scale proportional to the decay time $\tau$ for a sheared configuration to relax to zero energy after the driving strain is turned off.  
Recently, a direct calculation \cite{HIkeda1} of $\tau$  from the dynamical matrix of jammed configurations was found to give the same relationship between $\tau$ and $\theta$ as in \cite{DeGiuli,During}.  

If these marginal-stability arguments are correct (see  \cite{supp} for further discussion), and if the exponent $\theta$ has the same value in stress-isotropic jammed configurations as in stress-anisotropic jammed configurations, it could  imply a common universality for dynamic behavior.  Such a common value for $\theta$ 
was found for thermalized hard spheres at jamming in
 \cite{Jin2, Urbani}.  However it remains unclear whether the properties of the thermally equilibrated, mechanically stable,  shear-jammed states of  \cite{Jin2, Urbani} are necessarily the same as in the athermal, non-equilibrium, steady-state of shear-driven jamming.

{\color{black}Experimental support for the critical scaling of shear-driven flow curves in 3D has been found in both non-Brownian suspensions \cite{Nordstrom,Boyer} and  emulsions \cite{Paredes,Dinkgreve,Dinkgreve2}.  However, the critical exponents $\beta\approx 1.7 - 2$ found in these works are significantly smaller than that given by  the above theoretical prediction, $\beta=2.83$ \cite{supp}, possibly because the data used in these experiments span  too wide a range of packing $\phi$.  We are unaware of any similar experimental investigations for the divergence of relaxation times or  bulk viscosity in {\em athermal} compression-driven systems.}

Recently, {\color{black} numerical simulations have been used to investigate dynamic behavior below the jamming $\phi_J$.
As a direct probe of diverging time scales upon approaching jamming from below,}
Ikeda et al. \cite{Ikeda} measured the decay time $\tau$   as  3D configurations relax to zero energy according to overdamped equations of motion.
For both stress-isotropic random initial configurations,  and for stress-anisotropic initial configurations sampled from steady-state shearing,
they found  $\tau$ to collapse to a common curve, with a common divergence  as $\phi\to\phi_J$,  thus suggesting the same critical universality. 
However, a more recent work \cite{Nishikawa} by several of the same authors of \cite{Ikeda} questions these  results. While the predictions of \cite{DeGiuli,During,HIkeda1}, relating the divergence of $\tau$ to the force exponent $\theta$,
appear to hold for small system sizes, once the number of particles $N$ in the system is sufficiently large, they found that $\tau\sim\ln N$ for $\phi<\phi_J$; thus $\tau$ {\color{black}would seem to have} no proper thermodynamic limit.  
It is therefore important to re-examine  this question numerically, using a method alternative to $\tau$, to probe
the  time scale associated with jamming as $\phi\to\phi_J$ from below.

To do so, we  consider here isotropic compression at a finite rate $\edot$ \cite{Torq1} 
of soft-core, overdamped, athermal spheres, as in a non-Brownian suspension, in both   2D and 3D.  The finite rate $\edot$ introduces a control time  by which one can probe the  time scale associated with jamming.  Measuring the bulk viscosity $\zeta= p/\edot$, we find no finite-size effect, as was claimed for $\tau$ in \cite{Nishikawa}. Considering soft spheres allows us to measure not only  how $\zeta$  diverges below $\phi_J$, but also how $p$ behaves above $\phi_J$. 
We can then compare these results against previous simulations of the viscosity $\eta_p$ in the shear-driven steady-state.}

{\bf Model}:  Our model consists of bidisperse, frictionless, soft-core spheres, with equal numbers of big and small spheres with diameter ratio $d_b/d_s=1.4$ \cite{OHern}. For particles with center of mass positions $\mathbf{r}_i$, and $r_{ij}=|\mathbf{r}_i-\mathbf{r}_j|$, two particles interact with a one-sided harmonic contact potential, $U(r_{ij})=\frac{1}{2}k_e(1-r_{ij}/d_{ij})^2$, whenever their separation $r_{ij}<d_{ij}=(d_i+d_j)/2$.  The elastic force on $i$, due to contact with $j$, is thus $\mathbf{f}_{ij}^\mathrm{el}=-dU(r_{ij})/d\mathbf{r}_i$, and the total elastic force on $i$ is  $\mathbf{f}_i^\mathrm{el}=\sum_j\mathbf{f}_{ij}^\mathrm{el}$, where the sum is over all $j$ in contact with $i$.  Particles also experience a dissipative drag force  $\mathbf{f}^\mathrm{dis}_i$ with respect to a suspending host medium.  We take $\mathbf{f}_i^\mathrm{dis}=-k_dV_i[\mathbf{v}_i-\mathbf{v}_\mathrm{host}(\mathbf{r}_i)]$, where $V_i$ is the volume of particle $i$, and $\mathbf{v}_i=d\mathbf{r}_i/dt$.
For uniform compression we define the local velocity of the host medium as $\mathbf{v}_\mathrm{host}(\mathbf{r})=-\edot\mathbf{r}$.
This  simple model  has been widely used for sheared {\color{black}suspensions} \cite{OT1,OT2,OT3,Heussinger1,Lerner,Durian,Tewari,Andreotti,Vagberg.PRL.2014,DeGiuli,During,Berthier}.
Particles obey the equation of motion, 
$m_i[d\mathbf{v}_i/dt] = \mathbf{f}_i^\mathrm{el}+\mathbf{f}_i^\mathrm{dis}$, where $m_i$ is the mass of particle $i$, which we take proportional to its volume  $V_i$.

To simulate our model, we use dimensionless units of length, energy, and time so that $d_s=1$, $k_e=1$, and  $t_0=(D/2)k_dV_sd_s^2/k_e=1$, where $D=2$, 3 is the dimensionality of the system.  We define the quality factor $Q\equiv\tau_d/\tau_e=\sqrt{m_sk_e}/k_dV_sd_s$ as the ratio of the dissipative time  $\tau_d=m_s/(k_dV_s)$ and the elastic time $\tau_e=\sqrt{m_sd_s^2/k_e}$  \cite{Vag}.  
{\color{black}Note, $t_0=(D/2)\tau_e/Q$}.
We set the mass of the small particles $m_s$ so that $Q=0.01$ in 2D and 0.0225 in 3D, which puts our system in the strongly overdamped limit $Q<1$ 
{\color{black}where $p$ is independent of $Q$} \cite{Vag}.  We use  LAMMPS \cite{lammps}  to integrate the equations of motion, using a time step of $\Delta t/t_0=0.01$.  
Our system consists of $N$ particles in a cubic (square) box of length $L$.  We  compress by decreasing the box length at a fixed strain rate, $dL/dt=-\edot L$, while the particles are acted on by the compressing host medium via  $\mathbf{f}_i^\mathrm{dis}$.
This results in an increasing packing fraction $\phi=N(V_s+V_b)/(2L^D)$.  We take periodic boundary conditions in all directions. 
Compressing our system at rates from $\edot=10^{-5}$ down to $10^{-8.5}$, we measure the pressure $p$ of the elastic forces from the stress tensor $L^{-D}\sum_{i<j}\mathbf{f}_{ij}^\mathrm{el}\otimes(\mathbf{r}_i-\mathbf{r}_j)$, as a function of the packing $\phi$.  To check for finite-size effects, we compare systems with $N=16384$ and $N=32768$ particles, averaging over 10 independent random initial configurations for each size. Further details of our compression protocol can be found in our supplemental material \cite{supp}.

\begin{figure}
\centering
\includegraphics[width=3.2in]{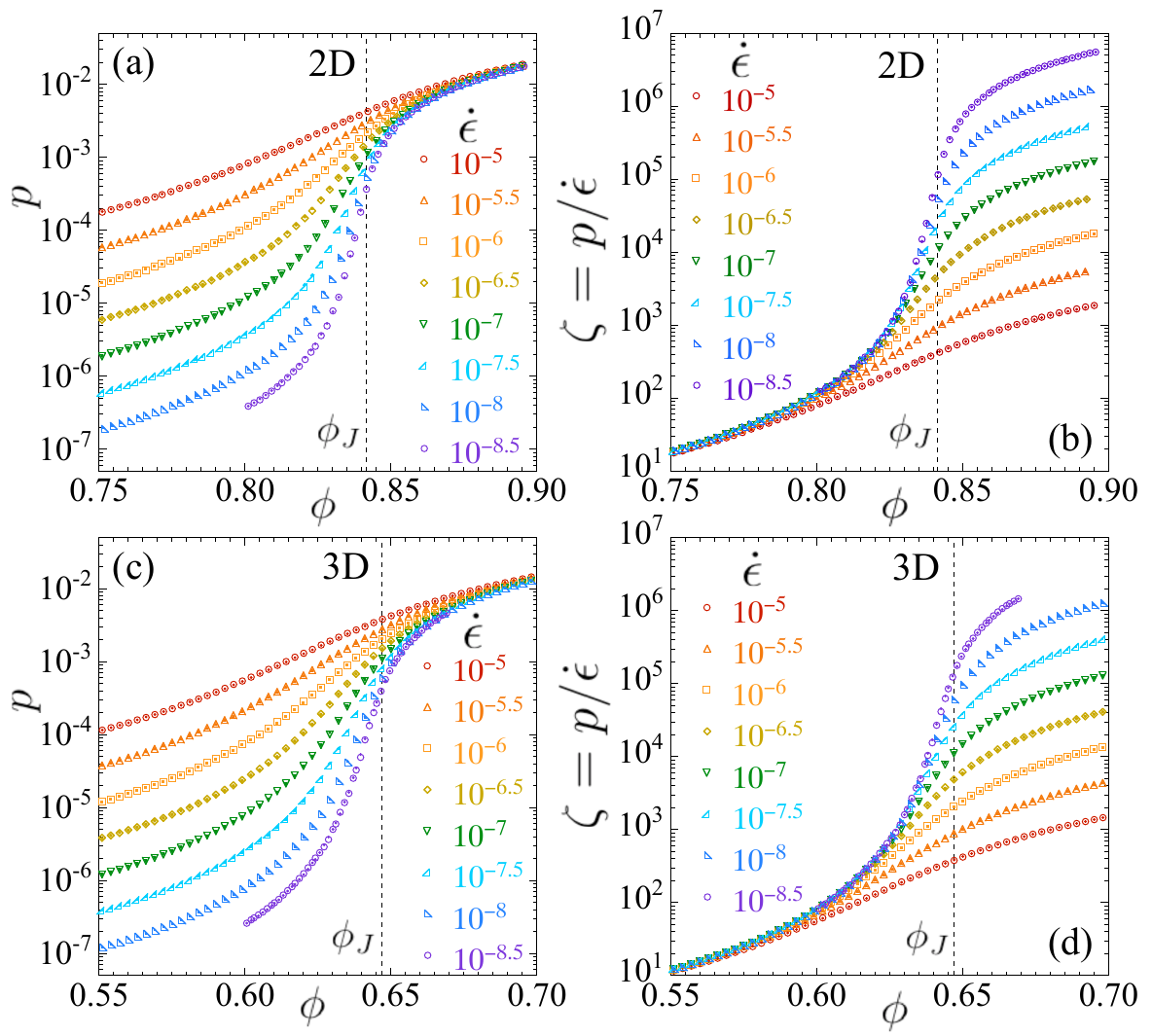}
\caption{(a) Pressure $p$  and (b) bulk viscosity $\zeta=p/\edot$ vs packing $\phi$,  for  different compression rates $\edot$ in two dimensions, and  (c) $p$ and (d) $\zeta$ in three dimensions.  The  vertical dashed lines locate the jamming $\phi_J$.  Results for $N=16384$ particles are shown as open symbols, while results for $N=32768$ are solid symbols.  No dependence on $N$ is observed.  Error bars are roughly the size of the data symbols. 
}
\label{p-zeta-vs-phi}
\end{figure}

{\bf Results}:  In Fig.~\ref{p-zeta-vs-phi} we plot our results for pressure $p$ and bulk viscosity $\zeta=p/\edot$ in both 2D and 3D.  No finite size effect is observed in our data (see  supplemental material \cite{supp} for  details). Our results are qualitatively similar to results seen for pressure and shear viscosity in shear-driven jamming \cite{OT2,VagbergOlssonTeitel}.  From the trends observed as $\edot$ decreases, our results 
suggest the following limiting behavior as $\edot\to 0$:  below $\phi_J$, $p$  vanishes  while $\zeta$ approaches a constant; above $\phi_J$, $p$ stays finite while $\zeta$ diverges. As $\phi\to\phi_J$ from above, $p$ vanishes continuously;  as $\phi\to\phi_J$ from below, $\zeta$ diverges continuously, demonstrating  the existence of a diverging time scale in compression-driven jamming. This is our first key result.

To confirm the above behavior, we posit that  pressure obeys a critical scaling equation of the same form found in shear-driven jamming  \cite{OT1,OT2,VagbergOlssonTeitel,OT3,Vagberg.PRL.2014},
\begin{equation}
p=\edot^q f(\delta\phi/\edot^{1/z\nu}),\qquad\delta\phi\equiv\phi-\phi_J
\label{escale}
\end{equation}
where $f(x)$ is an unknown scaling function.
Since we observe that $\zeta=p/\edot$ approaches a finite limit as $\edot\to 0$ below $\phi_J$, Eq.~(\ref{escale}) implies that $f(x\to-\infty)\sim |x|^{-(1-q)z\nu}$, so that for $\phi<\phi_J$,
\begin{equation}
\lim_{\edot\to 0}\,\zeta\sim |\phi-\phi_J|^{-\beta},\qquad\beta=(1-q)z\nu.
\label{ebeta}
\end{equation}
Above $\phi_J$, we observe that $p$ approaches a finite limit as $\edot\to 0$, so Eq.~(\ref{escale}) implies that $f(x\to+\infty)\sim x^{qz\nu}$, so that for $\phi<\phi_J$,
\begin{equation}
\lim_{\edot\to 0}\,p\sim(\phi-\phi_J)^y,\qquad y=qz\nu
\end{equation}
{\color{black}Note, the exponent $\beta$ is expected to be independent of the specific form of the elastic contact potential since it describes behavior in the $\edot\to 0$ hard-core limit \cite{OT3}; the exponent $y$, however, is sensitive to the power-law of the contact potential \cite{OHern,OT3}. A review of scaling in the context of shear-driven jamming may be found in \cite{VagbergOlssonTeitel}.}

Since we find no  size dependence in our data, we average the results from our $N=16384$ and $32768$ systems together,  so as to improve our statistics.  Expanding the log of the scaling function as a fifth-order polynomial, 
$\ln f(x)=\sum_{n=0}^5 c_nx^n$, we fit our data to Eq.~(\ref{escale}), regarding $\phi_J$, $q$, $1/z\nu$ and the $c_n$ as free fitting parameters.

The scaling form (\ref{escale}) holds only asymptotically close to the critical point, i.e., $\phi\to\phi_J$, $\edot\to 0$.  To test that our fits are stable and self consistent, we  fit to Eq.~(\ref{escale}) using different windows of data, with $\phi\in[\phi_\mathrm{min},\phi_\mathrm{max}]$ and $\edot\le\edot_\mathrm{max}$, to see how our fitted parameters vary as we shrink the data window closer to the critical point. Since our polynomial expansion for the scaling function $f(x)$ should be good only for small $x$, we also restrict the data used in the fit to satisfy $|x|\le 1$.

\begin{figure}
\centering
\includegraphics[width=3.2in]{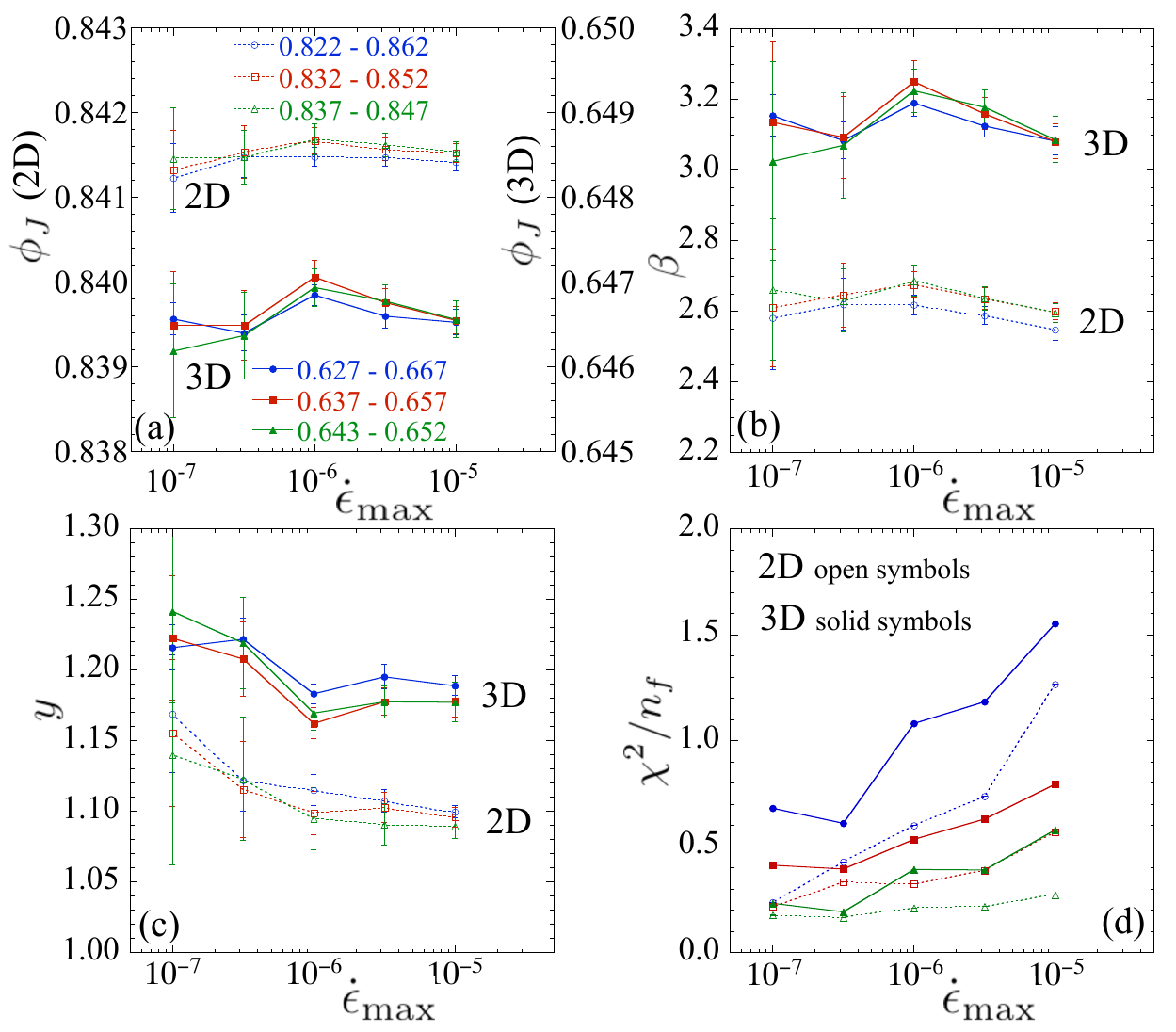}
\caption{Critical scaling parameters (a) $\phi_J$, (b) $\beta$,  (c) $y$, and (d) the $\chi^2/n_f$ of the fits, vs the upper limit of compression rate $\edot_\mathrm{max}$ used in the fit, for three different ranges of $\phi\in [\phi_\mathrm{min},\phi_\mathrm{max}]$.  Each panel shows results for both 2D and 3D systems.  We use the jackknife method to compute the estimated errors and bias-corrected averages of the fit parameters.  The data symbols in all panels follow the legend shown in (a); open symbols and dotted lines are for 2D, solid symbols and solid lines are for 3D. {\color{black}Note in (a) that the scale for $\phi_J$ in 2D is on the left, while the scale for $\phi_J$ in 3D is on the right.}
}
\label{fitparams-constphi}
\end{figure}

In Fig.~\ref{fitparams-constphi} we show the results from such fits, comparing  2D and 3D systems.  In Fig.~\ref{fitparams-constphi}(a) we show the jamming $\phi_J$, in \ref{fitparams-constphi}(b) the exponent $\beta$, in \ref{fitparams-constphi}(c) the exponent $y$, and in \ref{fitparams-constphi}(d) the $\chi^2/n_f$ of the fit, where $n_f$ is the number of degrees of freedom of the fit.  All quantities are plotted  vs  $\edot_\mathrm{max}$ for three different ranges of $[\phi_\mathrm{min},\phi_\mathrm{max}]$.
We use the jackknife method to estimate errors (one standard deviation  statistical error) and bias-corrected averages of these parameters.  
We see that the fitted parameters remain constant, within the estimated errors, as $\edot_\mathrm{max}$ decreases and we vary the range of $\phi$.  This suggests that our fits are  stable and self-consistent, with no need  to include corrections-to-scaling in the analysis, such as has been found to be necessary for simple shearing \cite{OT2,VagbergOlssonTeitel}.  The $\chi^2/n_f$ decrease as we narrow the window closer to the critical point; for our narrowest window in $\phi$ the $\chi^2/n_f$ remain roughly constant at the two smallest $\edot_\mathrm{max}$, another indication of the good quality of our fits.  It is difficult, however, to assess the significance of the numerical value of $\chi^2/n_f$; unlike for shearing, where each data point $(\phi,\gdot)$ represents an average over a steady-state shearing ensemble that is independent of its starting configuration \cite{Vagberg.PRE.2011}, for compression the configuration at a given $(\phi,\edot)$ is in general strongly correlated with the configuration at the previous compression step $(\phi-\Delta\phi,\edot)$, and so the estimated errors on the data points are similarly correlated.

\begin{figure}
\centering
\includegraphics[width=3.2in]{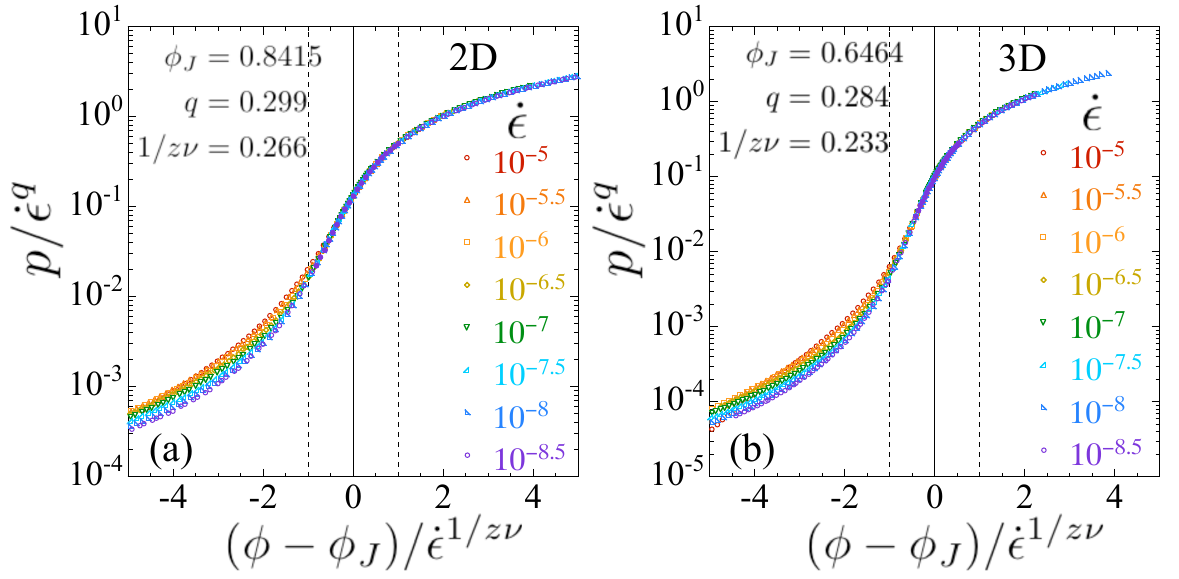}
\caption{Scaling collapses showing $p/\edot$ vs $(\phi-\phi_J)/\edot^{1/z\nu}$ for (a) our 2D system, and (b) our 3D system.  The values of $\phi_J$, $q$, and $1/z\nu$ used in making these plots come from our fits for $\edot_\mathrm{max}=10^{-6.5}$ and the narrowest range of $[\phi_\mathrm{min},\phi_\mathrm{max}]$.  The  points within this data window, that are used to make the fit, are shown as solid symbols; the points that are not used in the fit are shown as open symbols.  We see a good collapse even for data that lies well outside the data window used in the fit.  The vertical solid line locates the jamming $\delta\phi=0$; the vertical dashed lines denote the additional constraint $|x|\le 1$ for data used in the fit.
}
\label{pscaled}
\end{figure}

Fig.~\ref{fitparams-constphi} shows that the exponents $\beta$ and $y$ are different comparing 2D with 3D, in agreement with recent results for simple shearing \cite{Olsson3D}.  Thus  jamming criticality in 2D seems to be different from that in 3D. This is our second key result.  Taking the fit for the narrowest range $[\phi_\mathrm{min},\phi_\mathrm{max}]$ and  $\edot_\mathrm{max}=10^{-6.5}$ as representative, we use those parameters to make a scaling collapse of our data in Fig.~\ref{pscaled}, plotting $p/\edot^q$ versus $(\phi-\phi_J)/\edot^{1/z\nu}$.  We see  an excellent data collapse, which extends well outside the data window that was use to determine the fit parameters.  However, when $\delta\phi/\edot^{1/z\nu}\lesssim -2$, we see that the data  depart from a common scaling curve at the larger values of $\edot$.
We  believe this is due to the effect of corrections-to-scaling that become more significant as $\edot$ increases and one goes further from the critical point.

From the fits of Fig.~\ref{pscaled}
we find the following  critical parameters.  In 2D we have, $\phi_J=0.8415\pm 0.0003$, $\beta=2.63\pm 0.09$, and $y=1.12\pm 0.04$.  
We can compare these to the values found in simple shearing, in which case $\beta$ is the exponent associated with the divergence of the pressure analog of the shear viscosity, $\eta_p=p/\dot\gamma$.  For shearing of the same model system as considered here,  Ref.~\cite{OT2} gives $\phi_J=0.8435\pm0.0002$, $\beta=2.77\pm 0.20$, and $y=1.08\pm 0.03$, while  Ref.~\cite{OT3} gives $\phi_J=0.8433\pm 0.0001$, $\beta=2.58\pm0.10$, and $y=1.09\pm 0.01$.  We thus find that the values of the  exponents  $\beta$ and $y$, found here for compression-driven jamming, agree completely, within the estimated errors, with those found for simple shearing.  In 2D, compression-driven and shear-driven jamming appear to be in the same universality class.  This is our third key result.  

Note, our $\phi_J$ for compression-driven jamming is slightly lower than that found for shear-driven jamming.  It is well known \cite{Chaudhuri,Ciamarra,Vagberg.PRE.2011} that the value of $\phi_J$ can depend on the jamming protocol, and that the isotropic jamming $\phi_J$ found from rapid quenches of random initial configurations is lower than that found from shear-driven jamming.  
{\color{black}We can compare our $\phi_J$ for compression-driven jamming with previous values for isotropic rapid quenches.  In \cite{OHern2}, O'Hern et al. find $\phi_J=0.842$, while in \cite{VOT} V{\aa}berg et al. find $0.84177\pm 0.00001$.  Both agree, within the estimated errors, with our compression-driven value above.}

For our 3D system we find, $\phi_J=0.6464\pm 0.0005$, $\beta=3.07\pm0.15$, and $y=1.22\pm 0.03$.  
{\color{black}Our value of $\phi_J$ is a bit lower than the  $\phi_J=0.648$ found for the same model  with the rapid quench protocol \cite{OHern2}, and the  $\phi_J=0.6481$ found by Chaudhuri et al. \cite{Chaudhuri} for a more complicated isotropic compression/decompression protocol that starts at a low $\phi_\mathrm{init}$; neither of these works give an estimate for the error in their values.}  
As in 2D, our 3D compression-driven value of $\phi_J$ is slightly lower than values found for simple shearing  of the same model, $\phi_J=0.6474$ in \cite{Lerner} and \cite{Berthier}, and  $\phi_J=0.6491\pm 0.0001$ in \cite{Olsson3D}.

Concerning the critical exponents in 3D models of  overdamped sheared suspensions, 
numerical simulations on hard-core spheres by Lerner et al. \cite{Lerner} find $\beta=1/0.34=2.94$, while a later work of the same group, DeGiuli et al. \cite{DeGiuli}, find $\beta=1/0.36=2.8$.  Simulations on soft-core spheres by Kawaski et al. \cite{Berthier} find $\beta=1/0.391=2.56$.   
None of these works discuss the exponent $y$. 
More recent work by Olsson \cite{Olsson3D}, using   a  scaling analysis that includes corrections-to-scaling, finds $\beta=3.8\pm 0.1$ and $y=1.16\pm 0.01$. Olsson has argued that  other works find a smaller value of $\beta$ because they do not probe close enough to the critical point.  Given the disagreement among these values of $\beta$ for 3D simple shearing, our value of $\beta\approx 3.1$ for compression-driven jamming could be consistent with a common universality class.  The situation remains to be clarified.
{\color{black}See our supplemental material \cite{supp} for a comparison of $\beta$ with the marginal-stability  predictions.}

Note, the values of $y$ that we find from compression are {\color{black}in reasonable agreement with} the values found from shearing.   That $y>1$ for compression in both 2D and 3D is surprising since it has  generally been believed  \cite{OHern,Chaudhuri} that $y=1$ for our harmonic contact interaction.

The above results were obtained by averaging together independent runs at constant values of the packing $\phi$.  In our supplemental material \cite{supp} we repeat our scaling analysis, but averaging our runs at constant values of the average particle contact number $Z$.  We find no difference in any of the critical parameters between these two methods of averaging.

To summarize, we have carried out simulations of compression-driven jamming in a model of frictionless soft-core spheres in suspension, in two and three dimensions.  Using the compression rate $\edot$ as a scaling variable, in addition to the distance to jamming
 $\delta\phi$, we find that the pressure, and hence the bulk viscosity $\zeta$, obey a critical scaling law (\ref{escale}) of the same form as found previously for shear-driven jamming.  A diverging $\zeta$ demonstrates that compression is characterized by  a {\color{black}finite} time scale that diverges as $\phi\to\phi_J$ from below.  
Unlike the  claims in \cite{Nishikawa} for the relaxation time $\tau$, {\color{black}where  $\ln N$ finite size effects were seen for $\phi\le0.83$ in 2D  systems of size $N\ge4096$, and for $\phi\le0.57$ in 3D systems of size $N\ge1024$}, 
we observe no such finite size effects in the bulk viscosity $\zeta$ for the entire range of $\phi$ and $\edot$ we have used in our systems with $N=16384$  and 32768.
Our results indicate that isotropic, compression-driven, jamming in 2D and 3D have different critical exponents.  For 2D our results   suggest that stress-isotropic, compression-driven, jamming is in the same universality class as stress-anisotropic, shear-driven, jamming.  For 3D the situation is less clear, but our results could also be consistent with a common universality class.

\begin{acknowledgments}
We thank P. Olsson, T. A. Marschall,  M. A. Moore and {\color{black}H. Ikeda} for helpful discussions.
This work was supported by National Science Foundation Grant No. DMR-1809318. Computations were carried out at the Center for Integrated Research Computing at the University of Rochester. 
\end{acknowledgments}



\bibliographystyle{apsrev4-1}

\newpage

\begin{widetext}

\centerline{\bf{\large Critical Scaling of Compression-Driven Jamming of Athermal Frictionless Spheres in Suspension}}

\vskip 6pt
\centerline{\bf{\large Supplemental Material}}
\vskip 8pt
\centerline{Anton Peshkov and S. Teitel}
\centerline{\em Department of Physics and Astronomy, University of Rochester, Rochester, NY 14627}
\vskip 8pt
\end{widetext}

\setcounter{figure}{0}
\setcounter{equation}{0}
\setcounter{section}{0}
\renewcommand{\theequation}{SM-\arabic{equation}}
\renewcommand{\thefigure}{SM-\arabic{figure}}


\section{Compression Protocol}

To initiate our simulations, we start with configurations of randomly positioned particles at a small packing fraction $\phi_\mathrm{init}$.
To remove the large unphysical particle overlaps present in such configurations, we relax them towards a zero energy state, using our equations of motion without compression ($\edot=0$).  We then continue the simulations, compressing at a finite rate $\edot$ as described in the main text.
We find that the  pressure $p$, as one approaches the jamming $\phi_J$, is independent of the starting $\phi_\mathrm{init}$, provided $\phi_\mathrm{init}$ was taken sufficiently small.

\begin{figure}[h!]
\centering
\includegraphics[width=3.2in]{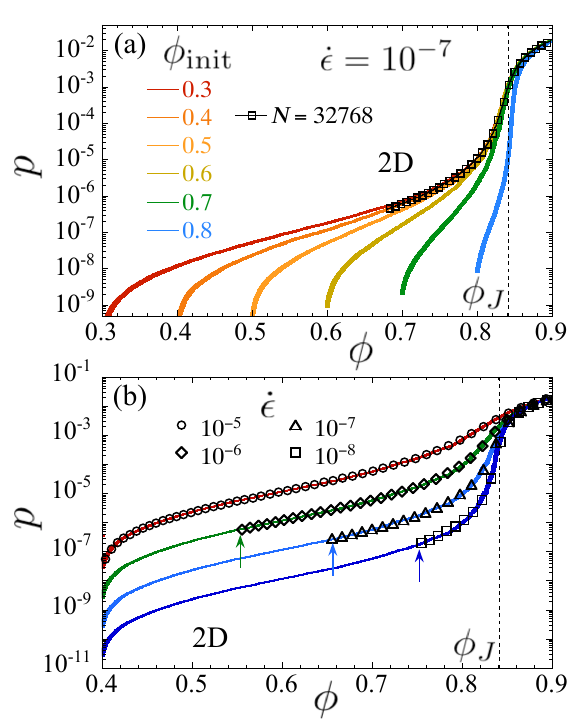}
\caption{(a) Pressure $p$ vs packing $\phi$ for a 2D system with $N=1024$ particles, at compression rate $\edot=10^{-7}$.  The different solid curves represent compression runs starting from different $\phi_\mathrm{init}=0.3-0.8$.  We see that the value of $p(\phi)$ becomes independent of $\phi_\mathrm{init}$ as $\phi$ increases towards jamming.  The open black squares  are results from the $N=32768$ system used in the main text.  We see perfect agreement with the smaller system size.
(b) Pressure $p$ vs packing $\phi$ for a 2D system compressed at rates $\edot=10^{-5}-10^{-8}$.  The solid curves are for a system with $N=1024$ particles, all starting from random configurations at the common $\phi_\mathrm{init}=0.4$.  The larger symbols represent data from the $N=16384$ system used in the main text, where compression starts from a larger $\phi^\prime_\mathrm{init}$ (indicated by the arrows) using a configuration from a run with a larger $\edot$.  We see perfect agreement between the two data sets.  In both panels, 
results are averaged over 10 independent initial configurations.  The width of each solid curve represents the estimated error.  Vertical dashed lines locate the jamming $\phi_J$.
}
\label{comp-test}
\end{figure}

To verify this, we performed test runs in 2D, starting with different values of $\phi_\mathrm{init}=0.3 - 0.8$, with a small system size of $N=1024$ particles compressed at a fixed rate $\edot=10^{-7}$.  In Fig.~\ref{comp-test}(a) we plot the resulting pressure $p$ vs packing $\phi$.
We see that $p(\phi)$ does depend on the value of $\phi_\mathrm{init}$ at the early stages of compression.  
However, as the system compresses and $\phi$ increases, the curves for different $\phi_\mathrm{init}$ approach a common limiting curve.  
Since, for our critical scaling analysis, we are only interested in behavior near jamming, for the 2D simulations described in the main text we chose $\phi_\mathrm{init}=0.4$.  Fig.~\ref{comp-test}(a) shows that this is small enough to remove all effects of the specific value of $\phi_\mathrm{init}$ on the values of $p$ for $\phi\gtrsim0.8$.  For our 3D system we chose $\phi_\mathrm{init}=0.2$.  Note, in Fig.~\ref{comp-test}(a) we also show as the open black squares our results for the $N=32768$ system that was used in the main text.  We see that these agree perfectly with the data from the smaller $N=1024$, indicating that the desired value of $\phi_\mathrm{init}$ does not depend on $N$.

For the large system sizes $N=16384$ and 32768 used in the main text, starting compression from the above small $\phi_\mathrm{init}$ becomes too time consuming at the slower compression rates, as one would spend much of the simulation time in the uninteresting region of low $\phi$.  To simulate more efficiently, we have adopted the following protocol.  For our largest compression rate $\edot=10^{-5}$ we compress from the small $\phi_\mathrm{init}$ as described above.  
For the next smaller rate, however, we initiate the simulation with a configuration taken
from the $\edot=10^{-5}$ run at some larger $\phi^\prime_\mathrm{init}>\phi_\mathrm{init}$.
This configuration is then compressed at the smaller rate $\edot$.
Provided $\phi^\prime_\mathrm{init}$ is  small enough  to be in the linear rheology regime where $p/\edot$ is independent of $\edot$, we find that the pressure in the initial configuration taken from the $\edot=10^{-5}$ run rapidly drops to the value appropriate for the smaller rate, and then follows a smooth curve that is independent of the value of $\phi^\prime_\mathrm{init}$.
We use the same algorithm for each successive $\edot$, initializing the run from the previous larger rate, using increasing values of $\phi^\prime_\mathrm{init}$ as $\edot$ decreases.

To validate this protocol, we performed test runs in 2D with $N=1024$ particles, compressing with rates $\edot=10^{-5} - 10^{-8}$, all starting from the same $\phi_\mathrm{init}=0.4$.
In Fig.~\ref{comp-test}(b) we plot (solid lines) the resulting pressure $p$ vs packing $\phi$.   On the same plot we indicate with larger symbols our results  from the $N=16384$ system used in the main text, where compression takes place using the above described protocol.  These latter runs are initiated at larger values of $\phi^\prime_\mathrm{init}$,  varying according to the value of $\edot$, as indicated by the arrows in the figure.  For all $\edot$ we find perfect agreement between these values of $p(\phi)$ and those of the smaller system that started from the common $\phi_\mathrm{init}$.

\section{Finite Size Effects}

Since we wish our analysis to be representative of behavior in the limit of an infinite sized system, it is important to demonstrate that our data do not suffer from effects due to the finite size of our numerical system.  Our results in Fig.~\ref{comp-test} clearly show that there are no finite size effects at small values of $\phi$.  However we still need to check that there are no finite size effects near jamming.
Since frictionless jamming is like a continuous phase transition with respect to the stress, one expects there to be a  correlation length $\xi$ that diverges as the critical point is approached.  Once one has $\xi \gtrsim L/2$, with $L$ the length of the numerical system, effects of the finite size of the system will appear.  We therefore want to check, for all our data points $(\phi,\edot)$, that our system is large enough  that they suffer from no such finite size effects.

\begin{figure}
\centering
\includegraphics[width=3.2in]{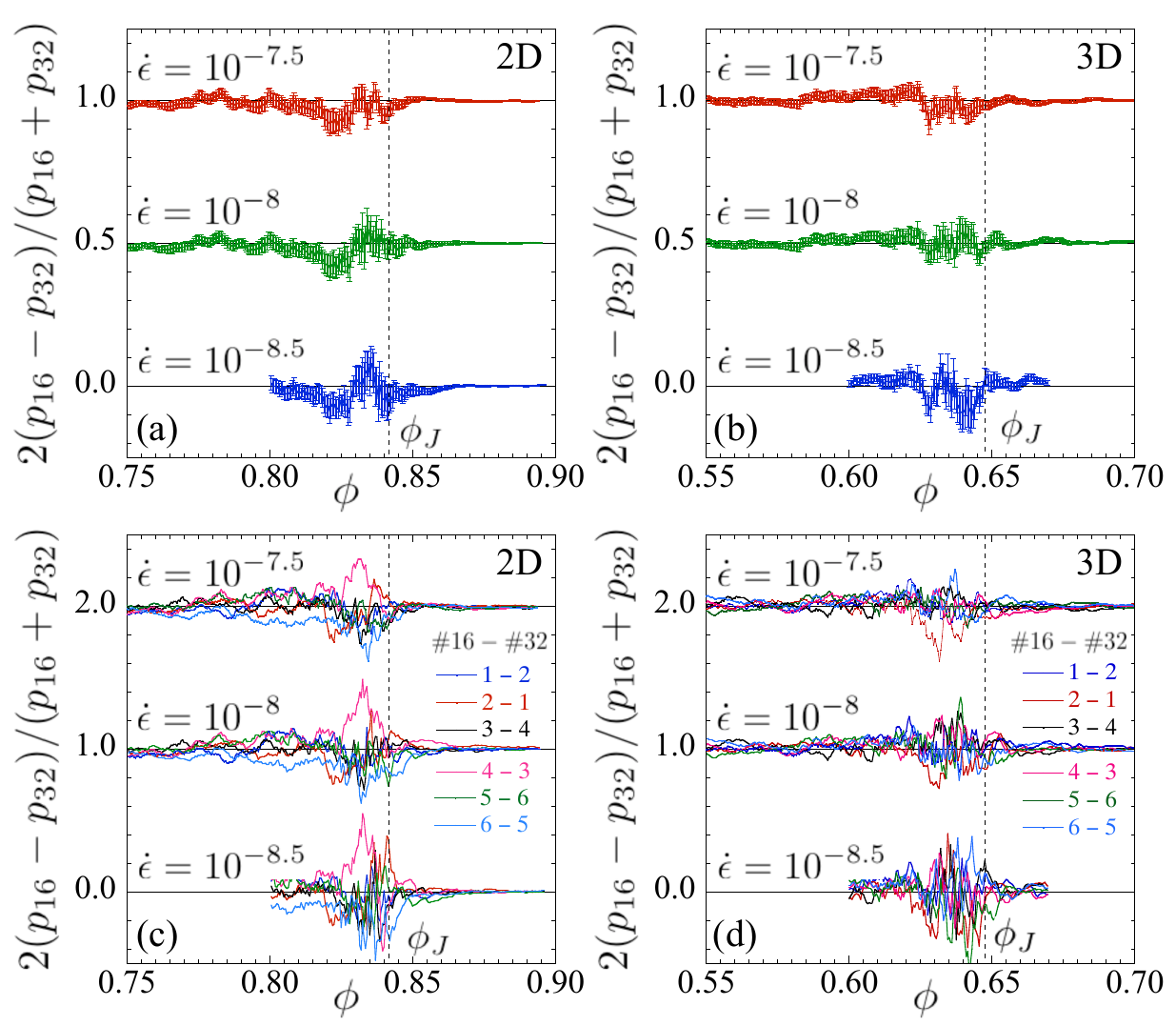}
\caption{Relative difference of pressure $p_{16}$ of a system with $N=16384$ particles compared to the pressure $p_{32}$ of a system with $N=32768$ particles.  We plot $\Delta p/p\equiv 2(p_{16}-p_{32})/(p_{16}+p_{32})$ vs $\phi$, showing results for our three smallest compression rates $\edot$.  Data for different $\edot$ are displaced vertically so as to easily distinguish the different data sets; all are fluctuating about zero. (a) and (b) show the difference in the pressure averaged over all 10 independent compression runs for 2D and 3D systems; (c) and (d) show the difference in pressure for six different pairs of individual samples from the two system sizes.  The legend ``\#16 -- \#32" indicates which sample from the smaller system is compared with which sample from the larger system.  The vertical dashed lines indicate the location of the jamming $\phi_J$.  }
\label{relE-p}
\end{figure}

Here we have simulated two different system sizes with $N=16384$ and $N=32768$ total particles.  We will denote the pressure in the first case $p_{16}$ and in the second case $p_{32}$.  In Fig.~\ref{relE-p} we plot the relative difference in pressure between these two different sized systems, $\Delta p/p\equiv 2(p_{16}-p_{32})/(p_{16}+p_{32})$, as a function of the particle packing $\phi$, for our three smallest compression rates $\edot$.  In Fig.~\ref{relE-p}(a) we show results for the average pressure (averaged over our 10 independent compression runs) for our 2D system; in \ref{relE-p}(b) we show the corresponding results for our 3D system.  The error bars represent one standard deviate of estimated statistical error.   Note, the results for each  $\edot$ are displaced vertically an amount 0.5 from the the next smaller $\edot$, so that one can easily distinguish the different data sets. One sees that $\Delta p/p$ fluctuates about zero, and all data points are within two standard deviations of zero.  This indicates that there are no systematic differences between the two system sizes, and that the finite $\Delta p/p$ is a consequence of statistical fluctuations in our finite sampling.  Not surprisingly, these fluctuations are largest when one gets close to $\phi_J$.

To further illustrate that the observed $\Delta p/p$ is due to statistical fluctuations and is not any systematic effect, in Figs.~\ref{relE-p}(c) and \ref{relE-p}(d) we plot $\Delta p/p$ for our 2D and 3D systems, but now computing the pressure difference between individual samples of the two system sizes, rather than the average over all samples.  We show results for six different pairs of samples.  The data for each $\edot$ is displaced vertically an amount 1.0 from the next smaller $\edot$, so that one can easily distinguish the different data sets.  Now we see that the sign of the fluctuation of $\Delta p/p$ about zero varies randomly  from one configuration pair to another.  Moreover, comparing the magnitude of the fluctuation of $\Delta p/p$ for the individual samples compared to the average over all samples, the latter is smaller by roughly the factor $1/\sqrt{N_s}$ (with $N_s=10$ the number of samples) that one would expect from statistical averaging.  We thus conclude that any difference we see comparing $p_{16}$ to $p_{32}$ is a statistical effect of finite sampling, rather than a systematic finite size effect.

We also note that we see no finite size effect in $p$, and hence in $\zeta=p/\edot$, even for the lower values of $\phi\lesssim 0.83$ in 2D, or $\phi\lesssim 0.57$ in 3D, where \cite{NishikawaS} reports finding a $\log N$ dependence of the decay time $\tau$ to relax to an unjammed state for systems of our size.  This is further illustrated in Fig.~\ref{comp-test}(b), where we compared our results  for $p$ in 2D for systems with $N=1024$ and $N=16384$ at lower $\phi$, and similarly see no finite size effects.
Thus, whatever is the dependence of $\tau$ on $N$, there does not seem to be any corresponding effect for $\zeta$.

\section{Exponents from the Marginal-Stability Analysis}

A key result of the infinite-dimensional mean-field theory of the jamming transition for thermalized hard-core spheres \cite{Charb2S,Charb1S} is that, exactly at $\phi_J$, the distribution of the magnitudes of the inter-particle contact forces $f_{ij}$ scales algebraically  as $f_{ij}\to 0$, $\mathcal{P}(f_{ij})\sim f_{ij}^\theta$, and that the exponent has the value $\theta=0.423$.  Numerical simulations of thermalized and athermal spheres in finite dimensions $d=2,3,4$ found values of $\theta$ consistent with this prediction, provided one excludes  contacts that are involved in only localized excitations of the system \cite{DeGiuli2S,Charb0S}.  It has been argued \cite{WyartS,Wyart3S,GoodrichS,CharbonneauS,Goodrich2S}  that the upper critical dimension for jamming may be $d=2$, and so mean-field critical exponents would apply in all dimensions $d>2$.

Using a marginal-stability analysis, the divergence of the pressure analog of shear viscosity $\eta_p=p/\gdot$ ($\dot\gamma$ is the shear strain rate) in the driven steady-state of a  uniformly sheared system has been argued \cite{DeGiuliS,DuringS} to be governed by this exponent $\theta$.  In \cite{LernerS} it was shown how $\eta_p$ is inversely proportional to the isolated smallest eigenvalue $\lambda_1$ of the dynamical matrix of the  configuration exactly at jamming.
More recently \cite{HIkeda1S} used a similar analysis to directly compute $\lambda_1$, and found the same relation to $\theta$.
In \cite{IkedaS} it was then numerically found that the relaxation time, for both an initially random and an initial sheared configuration to decay to an unjammed zero-energy configuration below $\phi_J$, followed the relation $\tau\sim 1/\lambda_1$.
These works thus imply $\eta_p\sim\tau\sim1/\lambda_1$.
The divergence of these quantities, as jamming is approached from below, can be stated in terms of the average contact number per particle $Z$, $\eta_p\sim\tau\sim\delta Z^{-\beta^\prime}$.  Here $\delta Z=Z_\mathrm{iso}-Z$, where $Z_\mathrm{iso}=2d$ is the isostatic value that occurs at jamming, and $Z$ is to be computed in the hard-core ($\dot\gamma\to 0$) limit after removing rattler particles.  Rattlers are particles which have unconstrained motion in at least one degree of freedom. 
In the marginal-stability calculations of \cite{DeGiuliS,DuringS,HIkeda1S}, the exponent $\beta^\prime$ is  related to  $\theta$  by $\beta^\prime=(4+2\theta)/(1+\theta)$.  Using $\theta=0.423$ one has $\beta^\prime=3.41$.

We choose to investigate critical behavior in terms of the packing fraction $\phi$ rather than $Z$, because $\phi$ is a directly controlled parameter, and because there is ambiguity how to define a rattler for soft-core particles driven out of equilibrium at finite strain rates $\edot$, such as we consider here (see more in the following section).  One needs to compute the hard-core, rattler free, value of $Z$ in order to apply the result $Z=Z_\mathrm{iso}$ at jamming, and so define $\delta Z$.  Viewing $\phi$ as the control parameter,  quantities diverge in the hard-core limit as $\eta_p\sim\tau\sim|\delta\phi|^{-\beta}$, and in \cite{DeGiuliS} a prediction is given that $\beta=(8+4\theta)/(3+\theta)=2.83$.   These two results then imply the relation $\delta Z\sim|\delta\phi|^{\beta/\beta^\prime}$ with $\beta/\beta^\prime=(2+2\theta)/(3+\theta)=0.83$, for $\phi\to\phi_J$ from below.


The hard-core limit is often defined in terms of  an infinite potential for particle overlaps.  However, for $\phi<\phi_J$, where energy relaxed configurations of even soft-core particles have no overlaps, the hard-core limit can also be taken as the quasi-static  limit for driven systems ($\dot\gamma\to 0$ for shearing, $\dot\epsilon\to 0$ for compressing), or the long-time limit of energy relaxing processes.
Simulations that have explicitly explored this hard-core limit have reported the following results.  Measuring $\eta_p$ for sheared  hard-core particles,  Lerner et al. \cite{LernerS} found in 2D $\beta^\prime=2.63$ and $\beta=2.17$, for $N=4096$  particles; in 3D  they found $\beta^\prime=2.94$ and $\beta=2.63$, for $N=2000$.  Similar simulations by DeGiuli et al. \cite{DeGiuliS} found $\beta^\prime=3.33$ and $\beta=2.78$ for  $N=1000$ in 3D.  Olsson measured the long time relaxation $\tau$ of $N=65538$  soft-core particles, relaxed to a zero energy configuration, using initial configurations sampled from steady-state shearing at a finite shear strain rate $\gdot$; in 2D he found $\beta^\prime=2.69$ and $\beta=2.71$ \cite{OlssonRelaxS}, while in 3D he found $\beta^\prime=3.7$ and (from analysis of $\eta_p$ rather than $\tau$) $\beta=3.8$ \cite{Olsson3DS}.
Most recently, Ikeda and Hukushima \cite{HIkeda2S} computed a quantity analogous to the bulk viscosity under quasi-static isotropic compression; using a finite-size scaling analysis for systems with $N\le 4096$ they claimed $\beta=1.9$ in 2D and 2.5 in 3D.

Ikeda et al. \cite{IkedaS}  measured  the long time relaxation $\tau$,  as well as explicitly computed the eigenvalue $\lambda_1$ of the energy relaxed configurations, for $N=3000$ particles in 3D.  For both initial random isotropic configurations and configurations sampled from shearing  at a finite $\gdot$, they found that all their data for $\lambda_1$ vs $\delta Z$ collapsed to a common curve with a $\beta^\prime=3.2$.  Nishikawa et al. \cite{NishikawaS}, however, repeated the calculation of $\tau$ for both isotropic and sheared initial configurations, but for much bigger system sizes up to $N=262144$.  For $N=4096$ they found $\beta^\prime=2.8$ in 2D and $\beta^\prime=3.3$ in 3D, similar to some of the previous results.  However as $N$ increased they found the surprising result that, for all $\phi<\phi_J$, $\tau$ grows $\sim \log N$  once $N$ is sufficiently large.  As one gets closer to $\phi_J$, one needs a larger $N$ to see this effect.  However they reported no such $\log N$ effect for $\eta_p$ of a sheared system.  As discussed in the previous sections, we see no such finite size effect in our measurement of pressure $p$, and hence the bulk viscosity $\zeta=p/\edot$, of our compressed system.

These  simulations  raise several questions concerning the application of the marginal-stability predictions to numerical results.  Nishikawa et al. \cite{NishikawaS} question  whether the long time relaxation $\tau$ is  a well defined quantity, and they conclude that ``the shear viscosity is finite in the thermodynamic limit, and that it decouples from the relaxation time at large $N$."  Thus viscosity may be more appropriate to consider than $\tau$.  The other simulations, using smaller systems which do not see such finite size effects, nevertheless still report a spread of values for $\beta$ and $\beta^\prime$.  It is hard to assess the accuracy of these results as the authors (except for Olsson) generally give few details about the fits that lead to the cited values.  As Olsson has noted \cite{Olsson3DS}, the fitted values of $\beta$ and $\beta^\prime$ tend to increase as one restricts the data used in the fitting to be closer to jamming.  Olsson's  analysis, with bigger system sizes than most others, also gives  evidence for $\beta=\beta^\prime$, in contrast to the prediction of \cite{DeGiuliS} that $\beta/\beta^\prime=0.83$. The conclusion $\delta Z\propto|\delta\phi|$, implied by $\beta=\beta^\prime$, was previously reported in simulations by Heussinger and Barat \cite{Heussinger1S}.  If correct,  the result $\beta=\beta^\prime$ would
raise  questions concerning the prediction of $\beta=2.83$ in \cite{DeGiuliS},  or whether $\beta=\beta^\prime$ reflects a more general breakdown of these theories in 2D and 3D.
These issues thus point to the need for further, careful, numerical simulations of shear and bulk viscosity; our current work is done with this motivation.

\section{Averaging at Constant  Contact Number $Z$}

In the main text of this paper we have averaged our independent samples together at constant values of the system packing $\phi$.  One may wonder if this is the best thing to do for the following considerations.  For a simple sheared system, when the system is sheared for a sufficiently long time, the average over the ensemble of sheared steady-state configurations becomes independent of the initial starting configuration \cite{Vagberg.PRE.2011S}.  Statistical fluctuations in the data at a given $(\phi,\dot\gamma)$ are thus, in principle,  independent of the fluctuations in the data at other $(\phi,\dot\gamma)$. 

\begin{figure}
\centering
\includegraphics[width=3.2in]{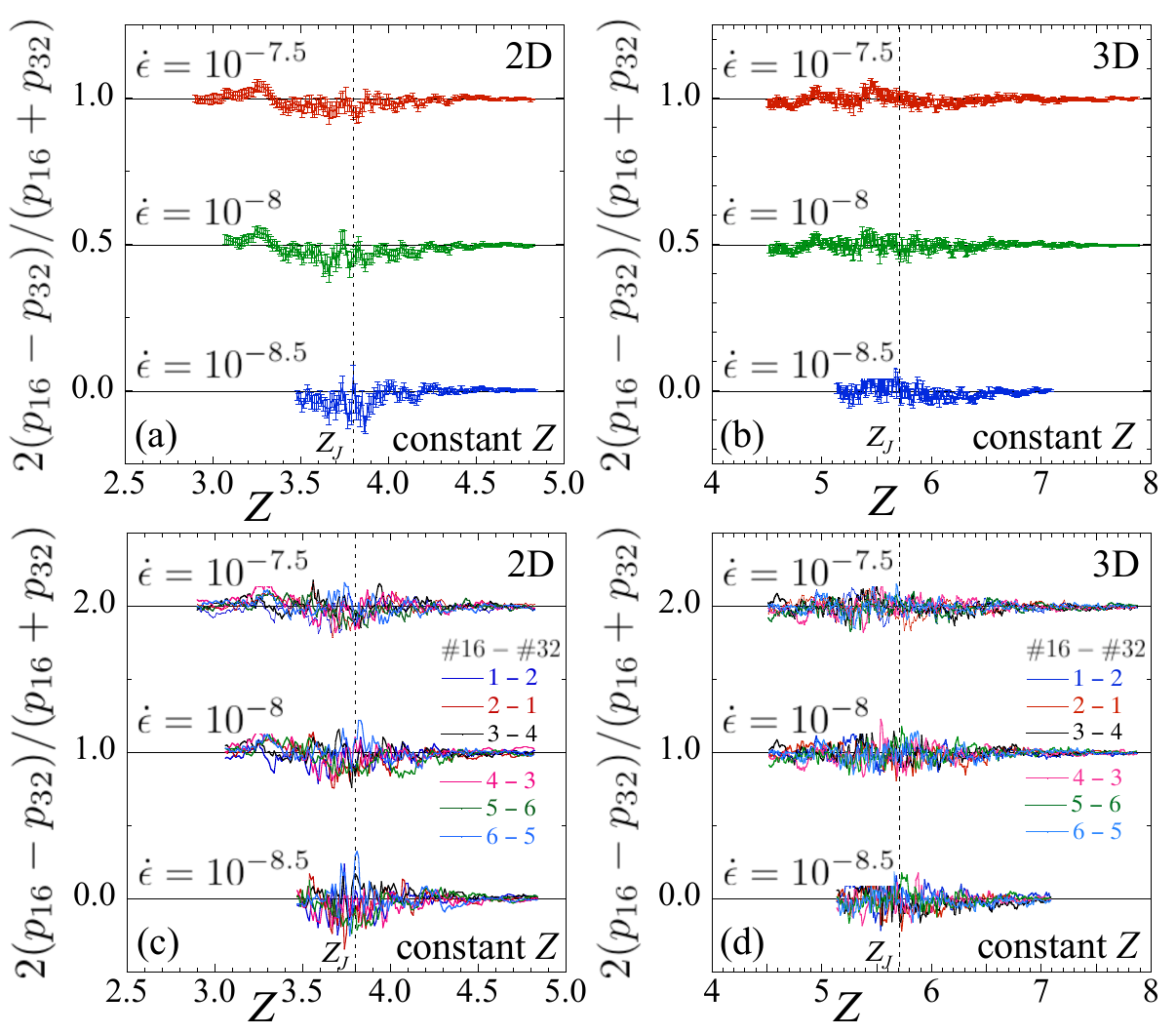}
\caption{Relative difference of pressure $p_{16}$ of a system with $N=16384$ particles compared to the pressure $p_{32}$ of a system with $N=32768$ particles.  We plot $\Delta p/p\equiv 2(p_{16}-p_{32})/(p_{16}+p_{32})$ vs $Z$, showing results for our three smallest compression rates $\edot$.  Data for different $\edot
$ are displaced vertically so as to easily distinguish the different data sets; all are fluctuating about zero. (a) and (b) show the difference in the pressure averaged at constant $Z$ over all 10 independent compression runs for 2D and 3D systems; (c) and (d) show the difference in pressure for six different pairs of individual samples from the two system sizes.  The legend ``\#16 -- \#32" indicates which sample from the smaller system is compared with which sample from the larger system.  The vertical dashed lines indicate the location of the jamming $Z_J$. }
\label{relE-p-constZ}
\end{figure}

For compression, however, the configuration at a given step $(\phi,\edot)$  is strongly correlated with the configuration at the previous step $(\phi-\Delta\phi,\edot)$.  It was found that the  jamming point  $\phi_{Ji}$, where a configuration first develops a finite pressure $p$, can depend on the particular  initial configuration $i$ from which the compression started \cite{OHernS,ChaudhuriS}.  For a  system with a finite number  of particles $N$, there will thus be a spread $\Delta\phi_J$ in these $\phi_{Ji}$.  This spread $\Delta\phi_{J}\to 0$ as $N\to \infty$  \cite{OHernS,VOTS}.  

\begin{figure}
\centering
\includegraphics[width=3.2in]{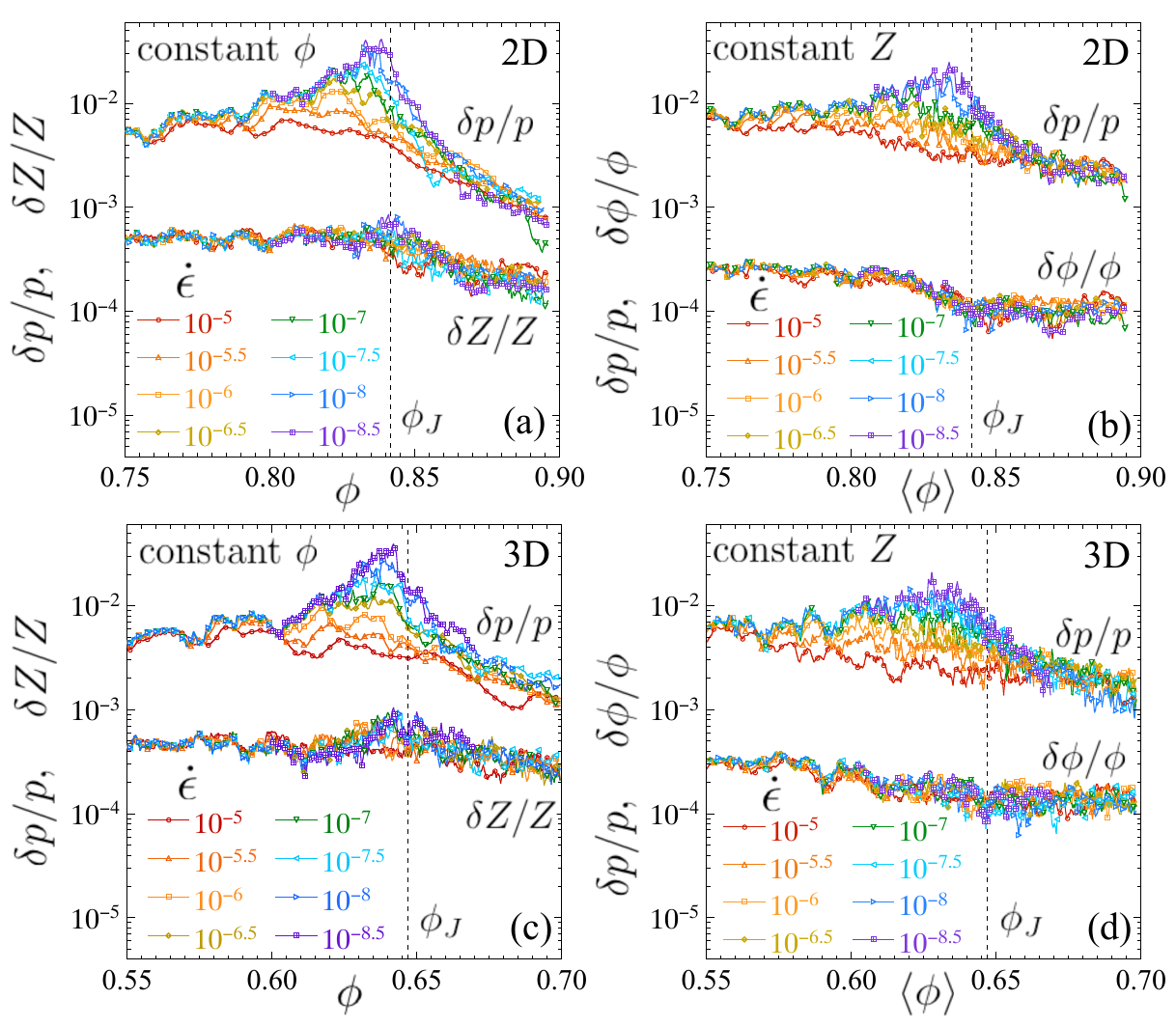}
\caption{Relative statistical errors in measured quantities for 2D and 3D systems, comparing averaging at constant $\phi$ with averaging at constant $Z$. (a) Relative errors  $\delta p/p$ and $\delta Z/Z$ vs $\phi$ for our 2D system, when averaging at constant $\phi$.  (b) Relative errors $\delta p/p$ and $\delta\phi/\phi$ vs the average $\langle\phi\rangle$ for our 2D system, when averaging at constant $Z$.  Results are shown for our different compression rates $\edot$.  (c) and (d) show the corresponding results for our 3D system.  Vertical dashed lines locate the jamming $\phi_J$.
}
\label{fluctuation}
\end{figure}

It is not clear if this behavior should affect the critical scaling analysis carried out in the main text of this work.  We are interested in the $\phi_J$ that characterizes the ensemble of compression runs, rather than any individual run.
We have found that the average  $\langle\phi_{Ji}\rangle $ is independent of the initial configurations, if these are taken randomly at sufficiently small $\phi_\mathrm{init}$.  However the width $\Delta\phi_{J}$ will be one source of fluctuation in the measured pressure, if averaging over configurations at constant $\phi$.  Because of this, several works  \cite{OHernS, ChaudhuriS} have analyzed critical properties by averaging configurations at constant values of $(\phi-\phi_{Ji})$, rather than constant $\phi$.

Alternatively, one could  average configurations at constant values of the average number of contacts per particle $Z$ \cite{LernerS,DeGiuliS,OlssonRelaxS}.  Even though different configurations of finite size systems may jam at different $\phi_{Ji}$, they all jam at the same isostatic contact number $Z_\mathrm{iso}=2d$, provided one has removed rattler particles \cite{OHernS} in the computation of $Z$. Here $d$ is the spatial dimensionality of the system.  Thus averaging at constant $Z$  removes the effect of the variations in $\phi_{Ji}$.  Because the identification of rattlers is most easily accomplished for mechanically stable configurations above jamming, and our configurations are dynamically generated, and so not in mechanical equilibrium, and also include configurations below $\phi_J$, we will not attempt to remove rattlers but rather we will compute the  contact number $Z$ averaged over all particles.  Thus our $Z_J$ at jamming will be slightly smaller than $Z_\mathrm{iso}$.   Nevertheless we can expect that averaging at constant $Z$ will still compensate for the variations in $\phi_{Ji}$, as there should on average be a fixed fraction of rattlers at jamming.  In this section we therefore repeat our analysis of the critical behavior of compression-driven jamming, but averaging our independent compression runs together at constant values of $Z$.  In the end we will find no differences in any of the critical parameters from those found in the main text, where we  averaged at constant $\phi$.

First we investigate  whether there are any finite size effects in our data, as we did for constant $\phi$ averaging, comparing systems of size $N=16384$ and $N=32768$.  Computing $\Delta p/p$, now averaging at constant $Z$,  we show our results in Fig.~\ref{relE-p-constZ}.   We again see no systematic finite size effects; the observed $\Delta p/p$ is consistent with the statistical effect of finite sampling.  Comparing with Fig.~\ref{relE-p} it appears that these statistical fluctuations are somewhat smaller when averaging at constant $Z$ as compared to averaging at constant $\phi$, particularly near jamming.  Since we find no evidence for any systematic finite size effect, in the analysis below we combine our results from the two system sizes so as to have 20 independent samples.

\begin{figure}
\centering
\includegraphics[width=3.2in]{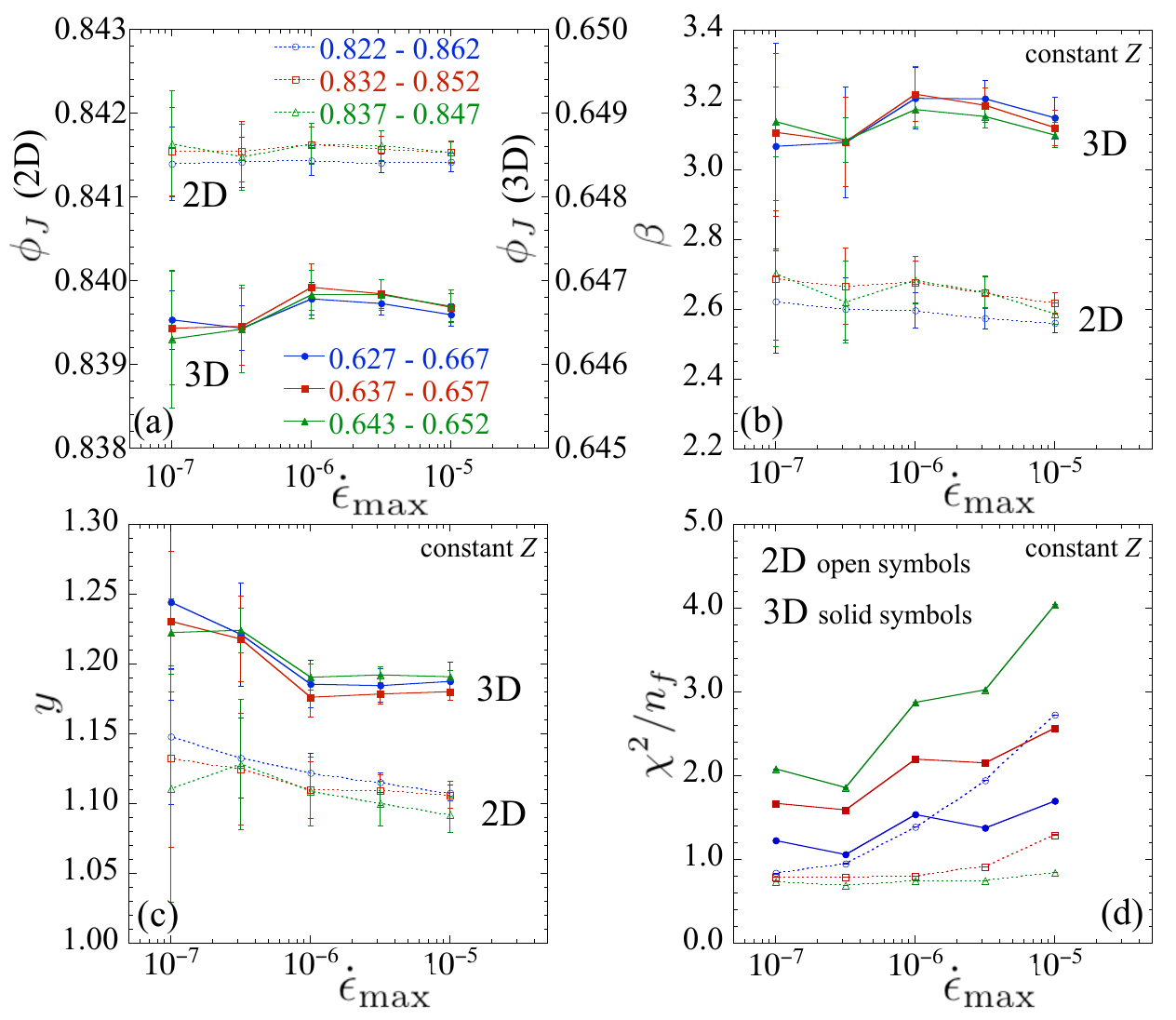}
\caption{Critical scaling parameters (a) $\phi_J$, (b) $\beta$,  (c) $y$, and (d) the $\chi^2/n_f$ of the fits, vs the upper limit of compression rate $\edot_\mathrm{max}$ used in the fit, for three different ranges of $\phi\in [\phi_\mathrm{min},\phi_\mathrm{max}]$.  Each panel shows results for both 2D and 3D systems.  We use the jackknife method to compute the estimated errors and bias-corrected averages of the fit parameters.  The data symbols in all panels follow the legend shown in (a); open symbols and dotted lines are for 2D, solid symbols and solid lines are for 3D.  The results shown here come from fits to our data when we have averaged our independent compression runs at constant values of the average contact number per particle $Z$. Note in (a) that the scale for $\phi_J$ in 2D is on the left, while the scale for $\phi_J$ in 3D is on the right.
}
\label{fitparams-constZ}
\end{figure}

Next we consider the relative statistical errors in the measured quantities, comparing averaging at constant $\phi$ with averaging at constant $Z$.  Since our compression runs are independent of one another, the estimated statistical error in pressure $\delta p$ is related to the standard deviation $\sigma_p$ of the distribution of pressures by $\delta p = \sigma_p/\sqrt{N_s}$, where $N_s=20$ is the number of samples.  
In Fig.~\ref{fluctuation}(a) we show the relative errors $\delta p/p$ and $\delta Z/Z$ vs the packing $\phi$ in our 2D system, for the case where we average our configurations together at constant $\phi$.  We show results for our different compression rates $\edot$.
In Fig.~\ref{fluctuation}(b) we similarly show $\delta p/p$ and $\delta \phi/\phi$ when we average at constant $Z$.  To make for an easier comparison, we plot these vs the average packing $\langle\phi\rangle$ rather than the fixed $Z$.
In Fig.~\ref{fluctuation}(c) and \ref{fluctuation}(d) we show  the same quantities for our 3D system.  

Not surprisingly, we see that the errors are largest near jamming.  The errors  $\delta p/p$ show a stronger variation with $\edot$, becoming larger as $\edot$ decreases,  than do the errors  $\delta Z/Z$ or $\delta \phi/\phi$, which are an order of magnitude or more smaller.  Comparing averaging at constant $\phi$ to averaging at constant $Z$, we see that the errors in the latter case are slightly smaller near $\phi_J$, as might be expected from the  discussion that introduced this section.  Note, however, that as we go either below or above $\phi_J$, the errors when we average at constant $Z$ become slightly larger than when we average at constant $\phi$.

The reduced fluctuations   between system sizes near $\phi_J$ seen in Fig.~\ref{relE-p-constZ}, and the reduced errors near $\phi_J$ seen in Fig.~\ref{fluctuation}, when we average at constant $Z$ as compared to constant $\phi$, suggest that averaging at constant $Z$ might give improved results for our scaling analysis.  However we find that this is not the case.  Using our values of $p$ and $\phi$, averaged over the different compression runs at constant $Z$, we fit to the scaling equation (1) of the main text using the same methods as described there.  In Fig.~\ref{fitparams-constZ} we show our results.  

In Fig.~\ref{fitparams-constZ}(a) we show the jamming $\phi_J$, in \ref{fitparams-constZ}(b) the exponent $\beta$, in \ref{fitparams-constZ}(c) the exponent $y$, and in \ref{fitparams-constZ}(d) the $\chi^2/n_f$ of the fit, where $n_f$ is the number of degrees of freedom of the fit.  For all quantities we plot our results vs  $\edot_\mathrm{max}$ for three different ranges of $[\phi_\mathrm{min},\phi_\mathrm{max}]$.
Comparing these to Fig.~2 of the main text, no appreciable difference is seen.  The fits are stable and self-consistent as we vary the window of data used in the fit.  Using  $\edot\_\mathrm{max}=10^{-6.5}$ and the narrowest range of $[\phi_\mathrm{min},\phi_\mathrm{max}]$, we find the following results.  In 2D we have, $\phi_J= 0.8415\pm 0.0004$, $\beta=2.62\pm 0.12$, and $y=1.13\pm 0.05$.  In 3D we have, $\phi_J=0.6464\pm 0.0005$, $\beta=3.08\pm 0.16$ and $y=1.22\pm 0.04$.  These are exactly the same values, within the estimated errors, as we found in the main text when averaging at constant $\phi$.  Moreover, the estimated errors found here are roughly the same, and in some cases a bit bigger, than we found in the main text.  We conclude that, for our system sizes, there is no advantage in averaging at constant $Z$ as compared to the simpler averaging at constant $\phi$.

\bibliographystyle{apsrev4-1}

\end{document}